  \documentclass[12pt]{iopart}  
\usepackage{iopams}
\usepackage{graphicx}
\begin{document}
%\documentclass[12pt]{iopart}
%\usepackage{graphicx,amssymb}
%\begin{document}
%\begin{frontmatter}
\title[S.V. Godambe et al.]{Very High Energy $\gamma$-ray observations of Mrk 501 using TACTIC imaging $\gamma$-ray telescope during 2005-06}
\
\author {S. V. Godambe, R. C. Rannot, P. Chandra, K. K. Yadav, A. K. Tickoo, 
K. Venugopal, N. Bhatt, S. Bhattacharyya, K. Chanchalani, V. K. Dhar,       
H. C. Goyal, R. K. Kaul, M. Kothari, S. Kotwal, M.K. Koul, R. Koul,
B. S. Sahaynathan, M. Sharma, S. Thoudam}

\address {Astrophysical Sciences Division, Bhabha Atomic Research Centre, Trombay, Mumbai 400085, India} 
\ead{rcrannot@barc.gov.in}
\begin{abstract}
% Text of abstract
In this paper we report on the Markarian 501 results obtained  during our  TeV $\gamma$-ray observations from  March 11 to May 12, 2005 and February 28 to May 7, 2006 for 112.5 hours   with the TACTIC $\gamma$-ray telescope. During 2005 observations for 45.7 hours, the source was found to be in a low state and we have placed  an upper limit   of 4.62 $\times$ 10$^{-12}$ photons cm$^{-2}$ s$^{-1}$ at  3$\sigma$ level on the integrated TeV $\gamma$-ray flux above 1 TeV from the source direction.
However, during the 2006 observations for  66.8h, detailed  data analysis revealed the presence of a  TeV $\gamma$-ray signal from the source  with a statistical significance of 7.5$\sigma$  above  $E_{\gamma}\geq$ 1 TeV. The time averaged  differential energy spectrum of the source in the energy range 1-11 TeV is found to match well with the power law function of the form ($d\Phi/dE=f_0 E^{-\Gamma}$) with $f_0=(1.66\pm0.52)\times 10^{-11}cm^{-2}s^{-1}TeV^{-1}$ and  $\Gamma=2.80\pm0.27$.

\end{abstract}

\section{Introduction}
%\label{}\section{Introduction}
Markarian 501 (Mrk 501) is a BL Lacertae (BL Lac) object at \textit{z} = 0.034 and  was discovered as a $\gamma$-ray source at $E_{\gamma}\geq$ 350 GeV by the Whipple collaboration \cite{Quinn96} and subsequently confirmed by the HEGRA collaboration\cite{Bradbury97}.
In March 1997 the source went into a  highly variable and strong emission state with flux doubling times of less than 0.5 day \cite{Quinn99}.   The time period of the outburst coincided with the source visibility windows of several ground-based Imaging Atmospheric Cerenkov Telescopes (IACTs). Thus almost continuous monitoring of Mrk 501 in TeV $\gamma$-rays  was possible with several IACTs (Whipple, HEGRA, CAT, TACTIC) located in the Northern Hemisphere   and they detected    an integral flux of up to  roughly 10 times   that from the Crab Nebula above 1 TeV\cite{Protheroe97}.
All Sky Monitor (ASM) on board the \textit{Rossi X-ray Timing Explorer} (RXTE), also reported the high X-ray activity of the source which started in March 1997 and continued till October 1997 \cite{Catanese97}.  The EGRET observations did not result in any detection of the source in high-energy $\gamma$-ray range (30MeV -30GeV)\cite{Catanese97} in 1997, though a year earlier, the same instrument had detected Mrk 501 at 5.2 $\sigma$ level for E$\ge$500 MeV\cite {Kataoka99a}. There was an  indication that the average optical U-band flux was higher in the month of the peak $\gamma$-ray activity \cite{Quinn99}.  Observations of Mrk 501 during 1998-1999 following the major 1997 outburst phase, reveal a mean emission level of $\frac{1}{3}$ of the Crab Nebula flux at 1 TeV, a factor of 10 lower than  the 1997 outburst. The VHE spectrum measured during this period  is substantially softer (by 0.44 $\pm$ 0.1 in spectral index) than the 1997 time-averaged spectrum \cite{Aharonian01}.   
The CELESTE group  has also detected a  weak $\gamma$-ray signal  from Mrk 501 (3.4$ \sigma$) using a  low threshold system at Themis during their  2000 observations\cite{smith06}.
In 2003-04 HESS observation of Mrk 501 at  large-zenith angle range for 1.8 hours, resulted in a marginal detection (3.1 $\sigma$) of the source\cite{Hess2005}. The MAGIC group has reported detection of two flaring episodes from the source on June 30 and July 09, 2005 with flux values of \mbox{3.48$\pm$0.10} 
and  \mbox{3.12$\pm$0.12} Crab units respectively  \cite{magic07}.
Among a number of  interesting source related features this group has unveiled  are the  detection of the IC peak in the SED for the most active nights, flux variability increase with energy and an intriguing 4 minute shift in time (on July 9, 2005) of the higher energy flare compared to lower energies.
\\

\section{Experimental Setup}

The TACTIC  atmospheric Cherenkov telescope is located  at Mt. Abu (24.6$^\circ$ N, 72.7$^\circ$ E, 1300 m  asl), a hill station in Western India. It   uses a tessellated light-collector of  9.5 $m^2$ area which is configured as a quasi-parabolic surface, yielding a measured spot-size of 0.3$^\circ$ for on-axis parallel rays. The PC-controlled 2-axes drive system gives a pointing / tracking accuracy of better than  $\pm$3 arc-mins. The telescope deploys a 349-pixel, photomultiplier tube (PMT) based imaging camera with a uniform pixel resolution of $\sim$ 0.3$^{\circ}$ and a field of view of  $\sim$ 6$^\circ$ $\times$ 6$^\circ$ to record images of atmospheric Cherenkov events produced by an incoming VHE cosmic-ray particle or a $\gamma$-ray photon.  Present observations use the inner 225 pixels (15 $\times$ 15 matrix ) of the camera for the  event-trigger generation,  based on the NNP (Nearest Neighbour Pair) topological logic. The data acquisition and control system of the telescope \cite{Yadav04} has been designed around a network of PCs running the QNX  real-time operating system. The details of the telescope functioning are presented elsewhere \cite{koul07}. The telescope is sensitive to $\gamma$-rays above 1 TeV and can detect the Crab Nebula at 5$\sigma$ significance level in 25 hours of ON source observation.
\begin{table}[b]
\caption{Observation log for Mrk 501} \label{tab:obser}
\begin{center}
\begin{tabular}{|c|c|c|c|c|}
\hline 
\textbf{Year} &\textbf{Month} &\textbf{Observation} &\textbf{Total Observation} &\textbf{Selected Observation}\\
&       &\textbf{Dates} &\textbf{Time (h)} &\textbf{Time (h)}\\             
\hline 
2005 &Mar. &11, 13, 15, 17-20   &11.6 &8.5   \\
2005 &Apr.  &05-10    &17.0      &15.9       \\
2005 &May        &02, 05-12        &33.8 &21.6      \\ 
\hline
2006 &Feb.       &28   &2.0   &2.0    \\ 
2006 &Mar.       &1-8, 25-29, 30 &28.6    &28.6      \\ 
2006 &Apr.      &1-6, 27-30  &20.3   &20.3    \\ 
2006 &May       &01-07  &15.9   &15.9    \\ 
\hline 
%\caption{Observation log for Mrk 501}
\end{tabular}
\end{center}
\end{table}

\section{Mrk 501 Observations}

TeV $\gamma$-ray observations of Mrk 501 presented in this paper were made  with       the  TACTIC $\gamma$-ray telescope during  two periods (1) 11 March to 12 May 2005 for 22 nights and (2) 28 February to 07 May  2006 for 33 nights, with a  total observation time of 129.25 hours. The observations were carried out in tracking mode, where the source is tracked continuously without taking  off- source data \cite{Quinn96}. This observation mode  improves the chances of recording possible flaring activity from the source direction. Details of the  Mrk 501 observations with the TACTIC $\gamma-ray$ telescope during 2005 and 2006 are given in Table \ref{tab:obser}. Several standard data quality checks have been used to evaluate the overall system behaviour and the general quality of the recorded data. These include conformity of the prompt coincidence rates with the expected zenith angle trend, compatibility of the arrival times of prompt coincidence events with the Poissonian statistics and the steady behaviour of the chance coincidence rate with time. After applying these cuts, we have  selected good quality data sets of 45.7 and 66.8 hours for 2005 and 2006 observations respectively, details of which are  given in Table \ref{tab:obser}. 

\section{Data Analysis and Results}
\begin{table}[b]
\caption{Cut Values used for Analysis} \label{tab:cuts}
\begin{center}
\begin{tabular}{|c|c|}
\hline 
\textbf{Parameters}              &\textbf{Cut Values}\\
\hline 
Length                  &0.11$^\circ$$\leq$ L$\leq$ (0.235+0.0265*log(size))$^\circ$\\
\hline
Width                   &0.065$^\circ$$\leq$ W $\leq$ (0.085+0.01200*log(size))$^\circ$\\
\hline  
Distance                &0.5$^\circ$ $\leq$ D $\leq$ 1.27$^\circ$\\
\hline
Size                    &S $\geq$ 350\\
\hline
Alpha                   & $\alpha$ $\leq$ 18$^\circ$\\
\hline
\end{tabular} 
\end{center}
\end{table}
\subsection{Data Analysis Procedure}
Detailed  analysis of data recorded by an atmospheric Cherenkov telescope,  involves a number of steps including  filtering of  night sky background light, accounting for the differences in the relative gains of the PMTs, finding Cherenkov image boundaries, image parameterization,  event selection etc. We have characterized each Cherenkov image by using a moment analysis methodology given in\cite{Hillas85,Reynolds93}. The roughly elliptical shape of the image is described by the LENGTH and the WIDTH parameters and its location and orientation within the telescope field of view are given by DISTANCE  and ALPHA parameters respectively. We also determine the two highest amplitude   signals recorded by the PMTs (max1, max2) and the amount of light in the image (SIZE) along with FRAC2 which is  the addition of two largest amplitude signals recorded by PMTs divided by the image size \cite{Weekes89}. The standard Dynamic Supercuts \cite{Mohanty98} procedure is then used to separate $\gamma$-ray like images from the huge background of cosmic-rays. In this procedure, the image shape parameters LENGTH and WIDTH have been used  as a function of the image SIZE so that energy dependence of these parameters can be taken into account. The $\gamma$-ray selection criteria (Table \ref{tab:cuts}) used in this analysis have been obtained on the basis of dedicated Monte Carlo simulations carried out for the TACTIC telescope.
Here, it may be noted that the cuts used in the present work  are softer particularly in terms of size parameter as compared to those used in \cite{Yadav07}. This is necessitated  by the need   to compensate for the  seasonal effects on the  energy threshold  due to  sky conditions   at Mt. Abu. 
For several years we are    observing VHE sources with the TACTIC telescope and  find that the average  winter sky (from November to  February) is slightly  better in terms of  sky transparency as compared to that during summer months ( March to  May ).
The dynamic cuts applied in the present data analysis  have also been validated  by applying them to the Crab Nebula data collected by  the telescope.
The $\gamma$-ray signal extraction, from the recorded data on a particular TeV $\gamma$-ray  source using a single imaging telescope     is generally, presented  by plotting  the histogram of $\alpha$  parameter (defined as the angle between the major axis of the image and the line between the image centroid and camera center, when the source is aligned along the optical axis of the telescope and hence the source is always placed at the centre of the camera)  after applying the set of  image cuts listed  in Table \ref{tab:cuts}.  $\alpha$  parameter distribution is expected to be flat for the isotropic background of cosmic ray events, whereas for the $\gamma$-ray signal  events, the distribution is expected to show a peak at smaller $\alpha$ values. This range for the TACTIC  telescope is  $\alpha$ $\leq$ 18$^{\circ}$. The contribution of the background events is estimated from a reasonably flat $\alpha$ region  from  27$^{\circ}$ $\leq$ $\alpha$ $\leq$ 81$^{\circ}$. The number of $\gamma$-ray events is then calculated by subtracting the expected number of background events, calculated on the basis of the background region of the $\alpha$ distribution \cite{Catanese98}, from the $\gamma$-ray domain events. The reason for not including the $\alpha$ bin 18$^{\circ}$-27$^{\circ}$ in the background region is to ensure that the background level is not overestimated because of a possible spill over of $\gamma$-ray events  into  this bin. Also,  the last bin has been dropped to make sure that underestimation of background level does not take place because of the truncation of the Cherenkov images recorded at the boundary of the imaging camera \cite{Catanese98}. The significance of the excess events has been finally calculated by using the maximum likelihood ratio method of Li \& Ma \cite{LiMa83}. 

\subsection{Validation of data analysis procedure using Crab Nebula data}
\begin{figure}
\begin{center}
\includegraphics[width=9cm,angle=0,clip]{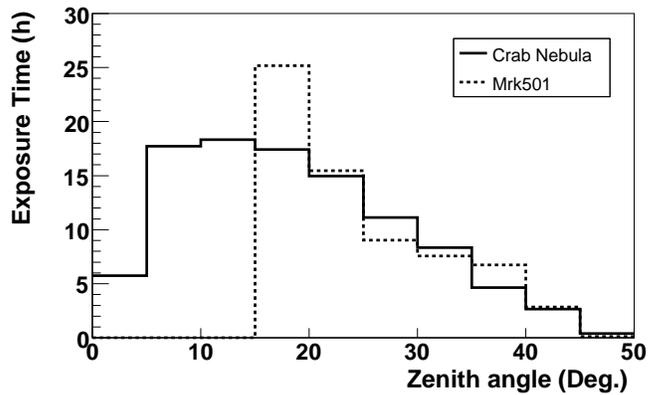}
\caption{Histogram coverage of the zenith angle range   from the Crab Nebula  direction (solid line) observed  for $\sim$101.4 hours during  Nov. 10, 2005 - Jan. 30, 2006 and  from the Mrk 501  direction (dotted line) observed  for 66.8 hours  during  Feb. 28, 2006 - May 7, 2006. }\label{fig:zen_range}
\end{center}
\end{figure}

\begin{figure}
\begin{center}
\includegraphics[width=14cm,angle=0,clip]{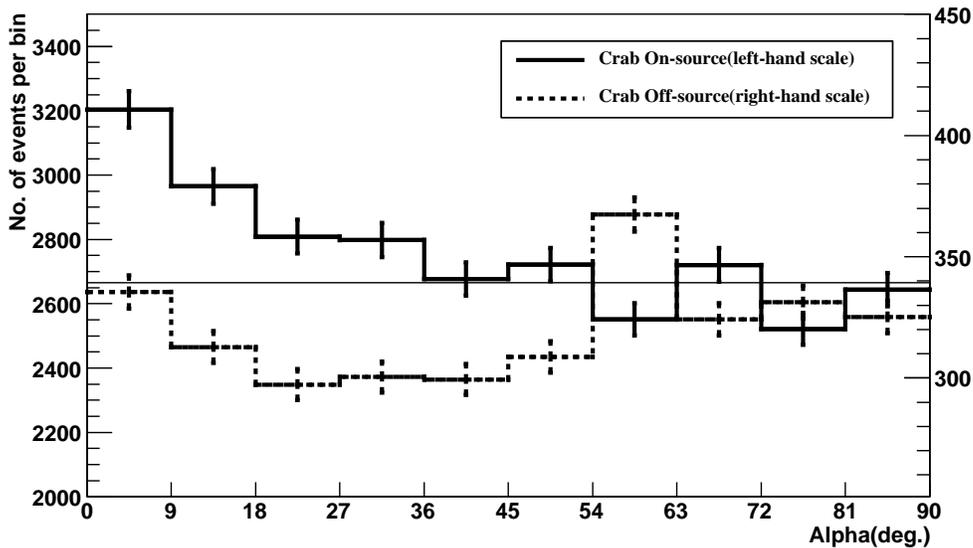}
\caption{ Distribution of image parameter ALPHA from the Crab Nebula  direction (solid line and  left-hand scale )   observed  for $\sim$101.4 hours during  Nov. 10, 2005 - Jan. 30, 2006. Horizontal line represents the background level per 9$^{\circ}$ bin derived using  reasonably flat $\alpha$ region  from  27$^{\circ}$ $\leq$ $\alpha$ $\leq$ 81$^{\circ}$.  Distribution of image parameter ALPHA is also shown ( dotted  line and right-hand scale)  from the Crab Nebula off-source  direction observed  for $\sim$12.5 hours during 2006. Error bars shown  are for statistical errors only.}\label{fig:alpha_crab}
\end{center}
\end{figure}

In order to test the validity of the data analysis procedure and in particular, the energy estimation procedure, we have first analyzed the Crab Nebula data collected  with   the TACTIC  imaging telescope  
for $\sim$101.4h during  Nov 10, 2005 - Jan 30, 2006 with zenith range covered as shown in Figure \ref{fig:zen_range}. The corresponding $\alpha$ plot after using the Dynamic Supercuts procedure   shown in Figure \ref{fig:alpha_crab} clearly indicates significant excess of events in the first two bins having  $\alpha$ range of 0$^{\circ}$ - 18$^{\circ}$, which is the $\gamma$ -ray domain in the plot for the TACTIC system. In order to show that this distribution is  flat in the absence of VHE  $\gamma$ -ray source, we have also shown the off -source data (for 12.5 hours only)  $\alpha$ distribution in Figure \ref{fig:alpha_crab}  using the same procedure. The events selected after using the cuts mentioned earlier yield an excess of $\sim$(839$\pm$89) $\gamma$-ray events with a statistical significance of  $\sim$9.64$\sigma$.  Here, it may be noted that we have derived the  background events from the on -source   $\alpha$ plot while  estimating   $\gamma$ -ray signal excess as we do not have the  Crab off- source data  of comparable duration  with the same zenith angle coverage.

The  corresponding  average  $\gamma$-ray rate  turns out to be  
$\sim$(8.27$\pm$0.88)/h. The same data sample was  then  analyzed again after restricting the zenith angle of the observations to  15$^\circ$ - 45$^\circ$  (similar to the zenith angle range which  Mrk 501 would cover as shown in Figure \ref{fig:zen_range}) so that the resulting $\gamma$-ray rate  can be designated as a approximate  reference of 1 Crab Unit (CU)  while  interpreting  the Mrk 501  data. The analysis yielded  an excess of $\sim$ (598$\pm$69) $\gamma$-ray events in an observation time of $\sim$63.3h with a corresponding $\gamma$-ray rate of $\sim$(9.44$\pm$1.09)/h, thus leading to  the conversion:
1 CU $\approx$ (9.44$\pm$ 1.09)/h.  This increase in $\gamma$-ray rate is due to the  superior $\gamma$-ray acceptance of Dynamic Supercuts at higher zenith angles which overcompensates the decrease in the rate due to increase in the threshold energy  of the telescope.\\ 

%\begin{figure}
%\begin{center}
%\includegraphics[width=14cm,angle=0,clip]{fig_3.eps}
%\caption{Distribution of image parameter ALPHA from the Crab Nebula off-source  direction observed  for $\sim$12.5 hours during 2006. Error bars shown  are for statistical errors only}\label{fig:alpha_craboff}
%\end{center}
%\end{figure}

Further,  the energy estimation procedure has been validated by deriving the Crab Nebula source energy spectrum using  the above mentioned  data  of $\sim$101.4 hours. The methodology used in deriving this energy spectrum is based on the Artificial Neural Network (ANN) technique and has been described in a detailed manner in \cite{Yadav07}.

The differential photon flux  per energy bin  has been computed using the formula
%---------------------------------------------------------------------------------
\begin{equation}
\frac{d\Phi}{dE}(E_i)=\frac {\Delta N_i}{\Delta E_i \sum \limits_{j=1}^5 A_{i,j} \eta_{i,j} T_j}
\end{equation}
%-------------------------------------------------------------------------------------
where $\Delta N_i$ and $d\Phi(E_i)/dE$ are the number of events and the differential flux at energy $E_i$, measured in the ith  energy bin $\Delta E_i$ and over the zenith angle range of 0$^\circ$-45$^\circ$, respectively. $T_j$ is the observation time in the jth zenith angle bin with corresponding energy dependent effective area ($A_{i,j}$) and $\gamma$-ray acceptance ($\eta_{i,j}$). The 5 zenith angle bins (j=1-5) used are 0$^\circ$-10$^\circ$, 10$^\circ$-20$^\circ$, 20$^\circ$-30$^\circ$, 30$^\circ$-40$^\circ$  and 40$^\circ$-50$^\circ$ with  simulation data  available at  5$^\circ$,15$^\circ$, 25$^\circ$,35$^\circ$ and 45$^\circ$. The number of $\gamma$-ray events  ($\Delta N_i$)  in a particular  energy bin is  calculated  by subtracting the expected number of background events, calculated  on the basis of background  region, from the  $\gamma$-ray domain events. \\

\begin{figure}
\begin{center}
\includegraphics[width=14cm,angle=0,clip]{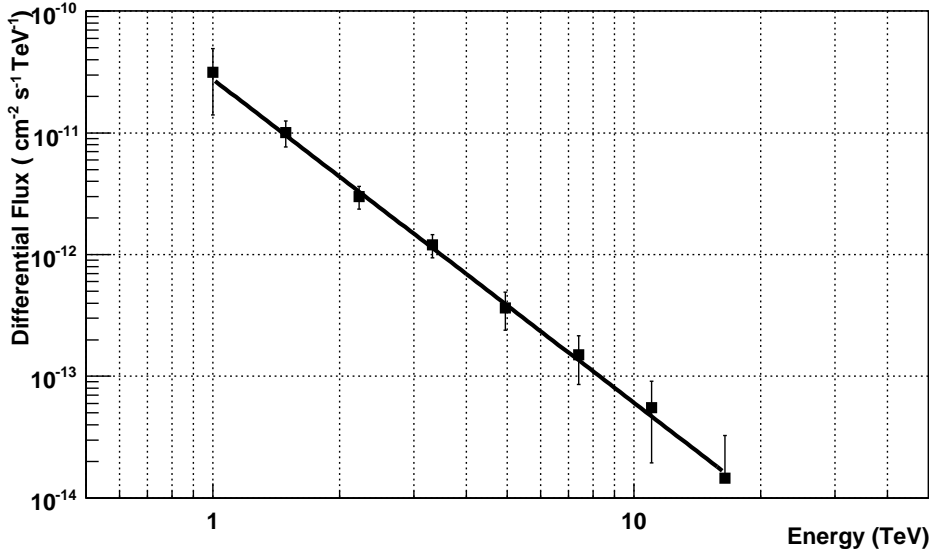}
\caption{Differential energy spectrum of Crab Nebula as measured by the TACTIC telescope during 2005-06. Error bars shown  are for statistical errors only.} \label{fig:crabsp}
\end{center}
\end{figure}

The $\gamma$-ray differential  spectrum  obtained   after applying the Dynamic Supercuts  and  appropriate values of  effective collection area and $\gamma$-ray acceptance  efficiency  (along with  their  energy and zenith angle dependence)  is shown in Figure \ref{fig:crabsp}.  Since  energy spectrum determination  also requires  an   'instrument calibration'  factor for converting image size in CDC counts to number of photoelectrons, this was determined by using the excess noise factor method. The analysis of relative calibration data yields  a value of  1pe $\cong$  (6.5$\pm$1.2) CDC for this  conversion factor when an average value of $\sim$1.7 is used for excess noise factor of the photomultiplier tubes.

\begin{table}
\caption{Crab Nebula differential energy spectrum data obtained during  2005-06 observations with the TACTIC telescope.}\label{tab:crabsp}
\begin{center}
\begin{tabular}{|c|c|c|c|}
\hline
\textbf{Energy}  &\textbf{Energy bin width }    &\textbf{Diff. flux  }    &\textbf{Statistical error in flux }\\
(TeV) & (TeV)   &   photons cm$^{-2}$ s$^{-1}$ TeV$^{-1}$ &  photons cm$^{-2}$ s$^{-1}$ TeV$^{-1}$ \\
\hline 
1.000   &0.403  &3.165 $\times$ 10$^{-11}$ &1.75 $\times$ 10$^{-11}$ \\
\hline
1.492   &0.601  &1.01 $\times$ 10$^{-11}$ &2.43 $\times$ 10$^{-12}$ \\
\hline
2.226   &0.896  &3.00 $\times$ 10$^{-12}$ &6.32 $\times$ 10$^{-13}$ \\
\hline
3.320   &1.337  &1.19 $\times$ 10$^{-12}$ &2.60 $\times$ 10$^{-13}$ \\
\hline
4.953   &1.994  &3.66$\times$ 10$^{-13}$ &1.26 $\times$ 10$^{-13}$ \\
\hline
7.389   &2.975  &1.5 $\times$ 10$^{-13}$ &6.5 $\times$ 10$^{-14}$ \\
\hline
11.023  &4.439  &5.54 $\times$ 10$^{-14}$ &3.59 $\times$ 10$^{-14}$ \\
\hline
16.445  &6.622  &1.45 $\times$ 10$^{-14}$ &1.79 $\times$ 10$^{-14}$ \\
\hline
\end{tabular} 
\end{center}
\end{table}

The  differential energy spectrum of the Crab Nebula  shown in Figure \ref{fig:crabsp} is of the form of  a  power law   $(d\Phi/dE=f_0 E^{-\Gamma})$  with  $f_0=(2.74\pm 0.64)\times 10^{-11} cm^{-2}s^{-1}TeV^{-1}$  and $\Gamma=2.65\pm 0.19$. The fit  has a $\chi^2/dof=0.53/6$ with a corresponding probability of 0.997. 
The low value of the reduced  $\chi^2$ may be due to overestimation of the errors      in the present measurements.
The errors in the flux constant and the spectral index are standard errors. The work on detailed   understanding of the telescope systematics is  in progress,
our preliminary estimates for the Crab Nebula spectrum indicate that the systematic
errors in flux and the spectral index are $\le$ $\pm$ 40 $\%$ and $\le$ $ \pm$ 0.42, respectively. Excellent matching of this spectrum with that obtained  by the Whipple and  HEGRA  groups  reassures   that the  procedure followed by us for obtaining the energy spectrum  of a $\gamma$-ray  source is quite  reliable.  Matching of the  Crab Nebula spectrum also  ensures  that   we have tested the full analysis chain  and the stability of the TACTIC  system directly  by using   $\gamma$-rays from the standard candle $\gamma$-ray source. The details of the derived Crab energy spectrum have  been tabulated in Table \ref{tab:crabsp}, wherein the energy binwidth is chosen in such a way that when it is divided by the corresponding energy it yields  a  constant value of 0.40.    \\

The data analysis procedure described above  was applied to the Mrk 501 datasets, obtained using the  TACTIC telescope during 2005-06. These datasets have been rigorously analysed,   on a yearly  as well as monthly observation spell basis,  for the presence of TeV $\gamma$-ray signal. 
We have also searched,   for  possible strong  TeV flaring episodes during  the epochs of our observations and  have further divided the data  on a nightly basis and repeated the  same analysis procedure. 

\subsection{Results of Mrk 501 2005 data  analysis } 

\begin{figure}
\begin{center}
\includegraphics[width=14cm,angle=0,clip]{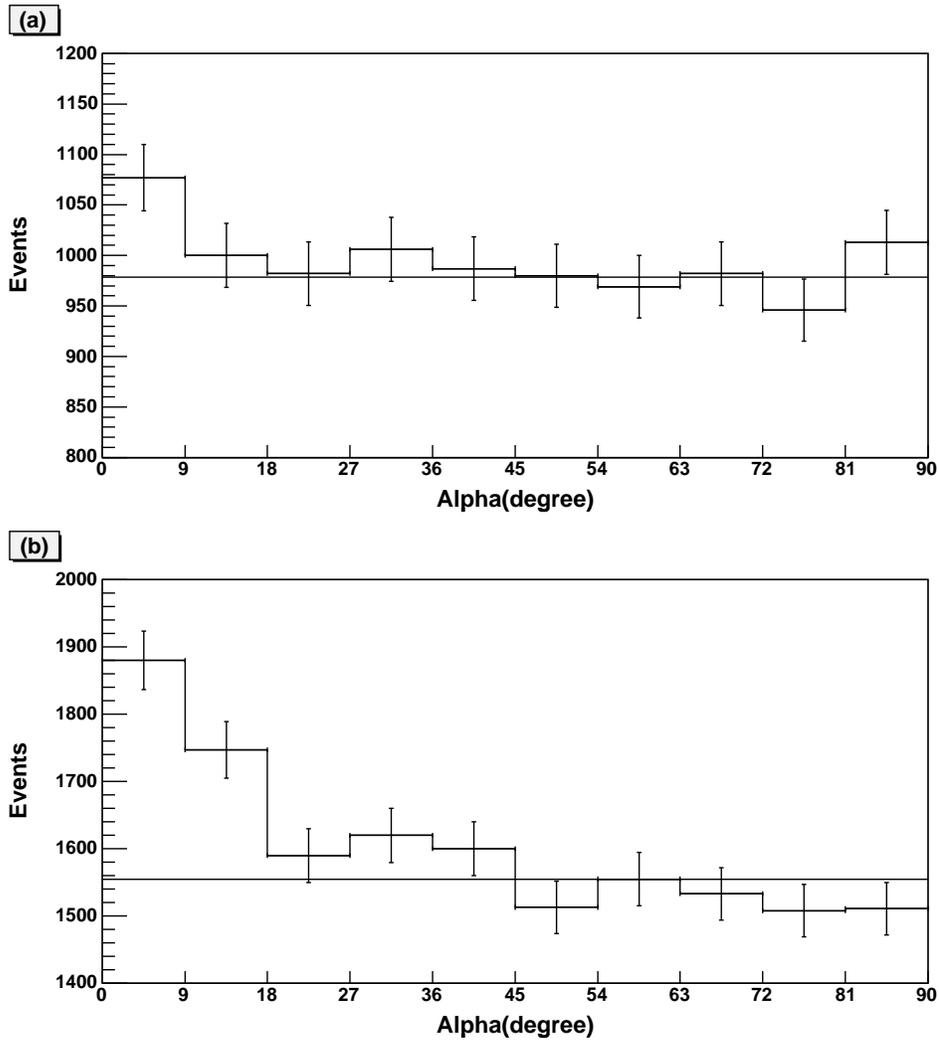}
\caption{Distribution of image parameter ALPHA from the Mrk 501 direction (a) 13 Mar. 2005 to 12 May 2005 (b) 28 Feb. 2006 to 07 May 2006. Horizontal line represents the background level per 9$^{\circ}$ bin derived using  reasonably flat $\alpha$ region  from  27$^{\circ}$ $\leq$ $\alpha$ $\leq$ 81$^{\circ}$. Error bars shown  are for statistical errors only.}\label{fig:alpha}
\end{center}
\end{figure}

When all the  data recorded during the year  2005 are analysed together, the corresponding results obtained are shown in Figure \ref{fig:alpha}a, wherein the histogram  of the alpha parameter has been plotted after having applied shape and orientation related imaging cuts given in Table \ref{tab:cuts}.  
As is clear from this Figure, the  distribution  is almost flat and the number of gamma-ray like events within $\gamma$- ray domain of the distribution are 120$\pm $ 52 with a statistical significance of 2.3$\sigma$.  
Thereby indicating that  the source was possibly  in a low TeV emission state (below TACTIC sensitivity level) during the period of these observations.
Moreover,  when 2005 data  are divided into three monthly spells I, II and III of March, April and May 2005 observations respectively and  by applying the same  data analysis procedure, the number of gamma-ray like events  obtained  are 8$\pm $ 25, 45$\pm $ 31 and 67$\pm $ 32 for  the respective spells.  These results which have  been tabulated in Table \ref{tab:month05} indicate  that the source TeV gamma-ray signal level has remained below TACTIC sensitivity   during  2005  observations.
  
\begin{figure}
\begin{center}

\includegraphics[width=14cm,angle=0,clip]{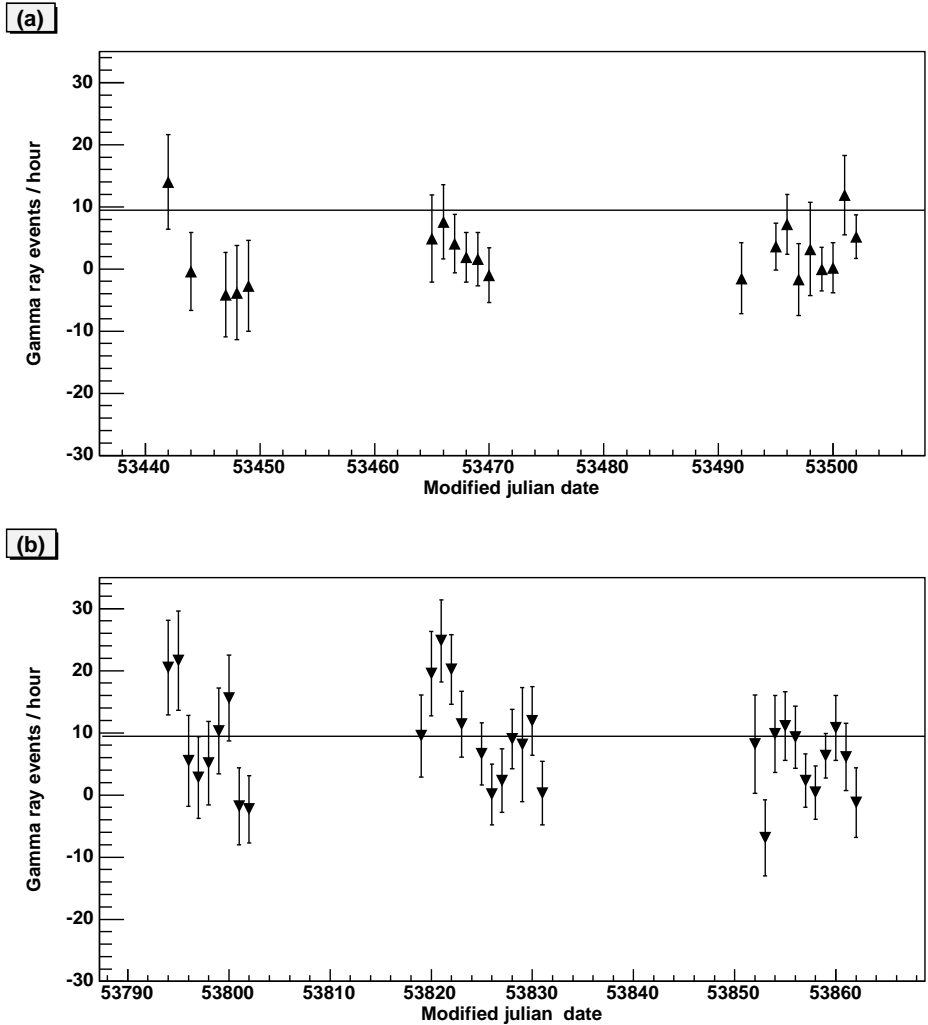}
\caption{TACTIC lightcurve of Mrk 501 for (a) 13 Mar. 2005 to 12 May 2005  (b) 28 Feb. 2006 to 07 May 2006. Horizontal line represents  the hourly   $\gamma$-ray rate as detected with the TACTIC telescope from the Crab Nebula direction. Points represent  nightly observed hourly   $\gamma$-ray rate from the Mrk 501 direction.  Error bars shown  are for statistical errors only.}\label{fig:lc}
\end{center}
\end{figure}

\begin{figure}
\begin{center}
\includegraphics[width=14cm,angle=0,clip]{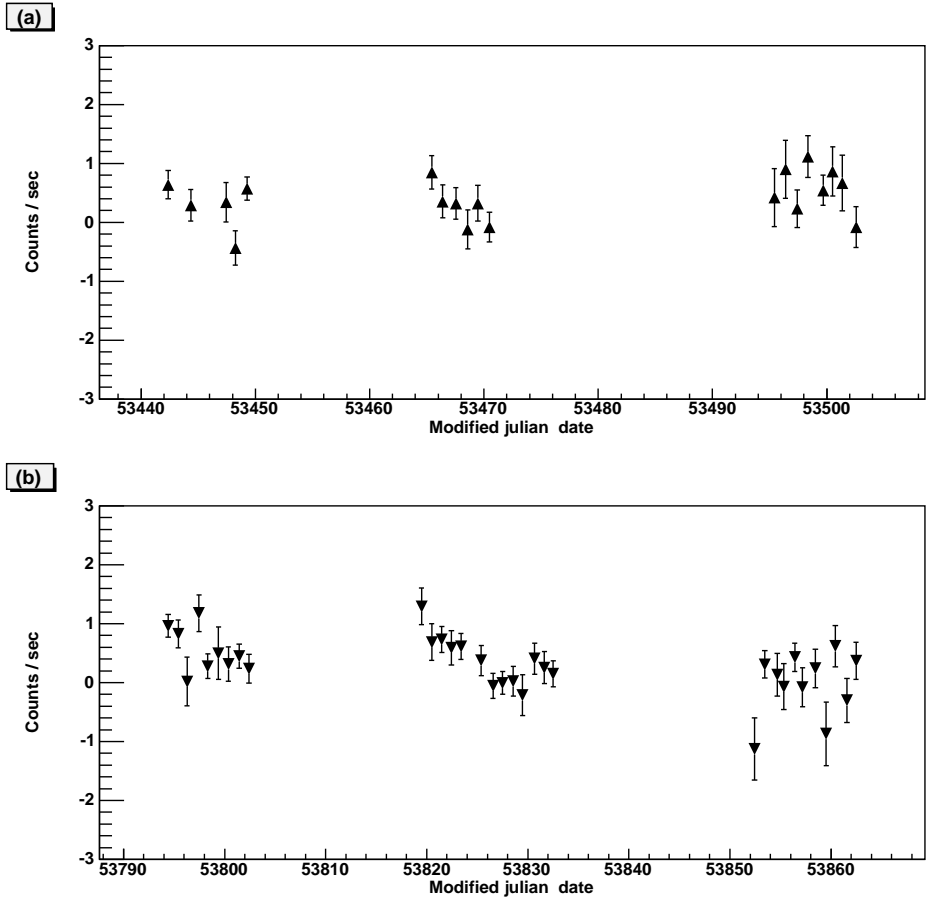}
\caption{ASM lightcurve of Mrk 501 for (a) March 13,  2005 to  May 12, 2005 (b)  February 28, 2006 to  May 7, 2006.}\label{fig:asm}
\end{center} 
\end{figure}

\begin{table}[h]
\caption{Monthly spell wise  analysis of Mrk 501 2005 data  with statistical  errors.}\label{tab:month05}
\begin{center}
\begin{tabular}{|c|c|c|c|}
\hline
\textbf{Spell}  &\textbf{$\gamma$-ray}  &\textbf{$\gamma$-ray} &\textbf{Significance ($\sigma$) } \\
\textbf{events}&\textbf{photons detected} &\textbf{rate/h} &\\
\hline
I   &8 $\pm$ 25 &1.02 $\pm$3.2  &0.32  \\
\hline
II    &45 $\pm$ 31 &2.81 $\pm$1.94  &1.4  \\
\hline
III    &67 $\pm$ 32 &3.1 $\pm$1.48  &2.0  \\
\hline
I+II+ III    &120 $\pm$ 52 &2.64 $\pm$1.14  &2.3  \\
\hline
\end{tabular}
\end{center}
\end{table}
Further, when 2005 data   are analysed on nightly basis, to explore the possibility of  very strong episodic TeV emission, the corresponding results obtained are depicted in Figure  \ref{fig:lc}a,  which shows the day-to-day variations of the $\gamma$-ray rate ($\gamma$-rays/hour) for 2005  observations. This light curve is characterised with  a reduced $\chi^2$ value of 11.99/19 with respect to the zero degree polynomial fitted  constant value   of 2.4 $\pm$ 1.1 photon events, with corresponding probability of  0.89 which is consistent with the no- variability hypothesis. 
The magnitude of an excess or deficit recorded on different nights is within $\pm$ 2 $\sigma$  level for 2005 observations and hence indicates the absence of a statistically significant episodic TeV gamma-ray signal. 
Further, in order to compare the VHE light curve with that of the source in the RXTE/ASM energy range 2-10 keV, we have used daily average count rates of ASM   from its archived data \cite{ASM},  to derive the source light curve for the contemporary period and the light curve so obtained is shown in Figure \ref{fig:asm}a. This light curve is characterised with  a reduced $\chi^2$ value of 27.8/18   with respect to the zero degree polynomial fitted constant value  of 0.40 $\pm$ 0.06 counts  and corresponding probability of 0.06, indicating consistency with the constant flux hypothesis and no variability   in a time scale of a day or more in the RXTE/ASM energy band also.  
We derive  an upper limit of 4.62 $\times$ 10$^{-12}$ photons cm$^{-2}$ s$^{-1}$ on the VHE $\gamma$-ray emission at 3$\sigma$ confidence  level using the total number of 120$\pm $ 52 gamma-ray like events obtained  during 2005 observations from the source direction  which  is about $28\%$ of the TACTIC detected integrated flux of the   Crab Nebula above  $E_{\gamma}\geq$ 1 TeV.  These results  together suggest that the source Mrk 501 was  possibly in a  low  state (below TACTIC sensitivity level)  during  2005  TACTIC observations. Here, we would like to mention  that the MAGIC observations on the same source during 2005 (May, 28 - July, 15 2005), wherein they have reported two episodes of strong flaring activity \cite{magic07},  do not overlap in time with the presented TACTIC observations which were taken between  March 11 -May 12, 2005. To some extent, TACTIC 2005  observations do provide the source light curve for epochs just  before the  MAGIC detected two  historical VHE flares    from Mrk 501   on June, 30 and July, 9  2005, which have  created a lot of excitement  in the field.      
 
\subsection{Results of Mrk 501 2006 data  analysis   }

%\subsection{Data analysis  results of Mrk 501 2006 data }
Similar  data analysis methodology was   used while analysing data recorded, using  TACTIC from the same source,   during    2006  observations.
When all the  2006 data  are analysed together, the corresponding results obtained are shown in Figure  \ref{fig:alpha}b, wherein again   a histogram  of the alpha parameter has been plotted  after having applied  imaging cuts given in Table \ref{tab:cuts}. This Figure shows that  the  distribution   is not  flat and the number of gamma-ray like events within the $\gamma$- ray domain of the distribution are 517 $\pm$ 69 with a statistical significance of 7.5$\sigma$.
Thereby indicating that  the source was   in a relatively high  TeV emission state as compared to that during our 2005  observations.

Next, we have divided 2006 data   into three monthly spells I, II and III of  February 28  to   March 8, 2006,  March 25 to April 7, 2006  and April 27  to May 7, 2006 observations respectively and   applied  the same  data analysis procedure for each monthly spell. The results of this exercise have been tabulated in Table \ref{tab:month06}.  The number of gamma-ray like events so obtained  are 147 $\pm$39, 261 $\pm$44 and 108 $\pm$ 33  with statistical significance of 3.7, 5.9 and 3.2 $\sigma$ for  each  spell respectively. This indicates that the  TeV gamma-ray emission level of the source has brightened   during  II spell  2006 observations as its statistical significance is close to 6$\sigma$ in 27.71 hours of observations.
\begin{table}[h]
\caption{ Monthly spell wise  analysis of Mrk 501 2006 data with only statistical  errors.  }\label{tab:month06}
\begin{center}
\begin{tabular}{|c|c|c|c|}
\hline
\textbf{Spell}  &\textbf{$\gamma$-ray}  &\textbf{$\gamma$-ray} &\textbf{Significance ($\sigma$) }\\
\textbf{events}&\textbf{photons detected} &\textbf{rate/h} &\\
\hline
I    &147 $\pm$39  &8.4 $\pm$2.2  &3.7  \\
\hline
II  &261 $\pm$ 44 &9.4 $\pm$1.5  &5.9  \\
\hline
III    &108 $\pm$ 33 &4.9 $\pm$1.5  &3.2  \\
\hline
I+ II+III    &517 $\pm$ 69 &7.7 $\pm$1.01  &7.5  \\
\hline
\end{tabular}
\end{center}
\end{table}
Further,  2006 data   were also  analysed on a nightly basis, to explore the possibility of  a  strong episodic TeV emission and  the corresponding results obtained are depicted in Figure \ref{fig:lc}b. This light curve is characterised by  a reduced $\chi^2$ value of 48.8/31 with respect 
to the zero degree polynomial fitted  constant value   
of 6.96 $\pm$ 0.99 counts and corresponding probability of 0.02.  This indicates some variability with respect to the constant flux hypothesis in a time scale of about 24 hours. The corresponding ASM light curve of the source is shown in Figure \ref{fig:asm}b \cite{ASM}. This light curve is characterised by  a reduced $\chi^2$ value of 69.5/31 with respect to the zero degree polynomial fitted  constant value  of 0.28 $\pm$ 0.05 counts and corresponding probability of 8.8$\times 10^{-5}$. Thereby indicating  variability  with respect to the constant flux hypothesis in a time scale of about 24 hours so  supporting TeV  observations in term of the type of variability mentioned above.        

Here, we would like to mention the  results of our observations during three nights of March 26, 27 and 28 , 2006 (MJD 53820, 53821 and  53822 ) when we find hourly gamma-ray rates of more than two times  that of the Crab Nebula. The excess of gamma-ray like events obtained are 37.0  $\pm$ 12.7,  41.7 $\pm$ 11.7 and   56.0 $\pm$ 15.5 for 1.9, 1.7 and 2.8 hours of observations  respectively. These  observations possibly  suggest that the source has  gone into a  highly active  state in the VHE region during these three nights. We are not in a position to study variability features of the source on a nightly basis in the TeV domain mainly due to the poor statistics obtained with the TACTIC telescope.

\subsection{Energy Spectrum of Mrk 501}

We have used the 2006 data of gamma-ray like events  to determine the source  differential energy spectrum. The spectrum extends up to the 11 TeV with power-law index of 2.80 $\pm$ 0.27. The details of the method used to derive the differential time-averaged energy spectrum of the Mrk 501 are given elsewhere \cite{Yadav07}. 

\begin{table}
\caption{Differential energy spectrum data for Mrk 501 in 2006 with the TACTIC telescope. Only statistical errors are  given below.}\label{tab:501sp}
\begin{center}
\begin{tabular}{|c|c|c|c|}
\hline
\textbf{Energy}  &\textbf{Energy bin width}    &\textbf{Diff. flux  }    &\textbf{Statistical error in flux }\\
(TeV) & (TeV)   &   photons cm$^{-2}$ s$^{-1}$TeV$^{-1}$ &  photons cm$^{-2}$ s$^{-1}$TeV$^{-1}$  \\
\hline 
1.000   &0.403  &1.57 $\times$ 10$^{-11}$ &1.28 $\times$ 10$^{-11}$ \\
\hline
1.492   &0.601  &6.31 $\times$ 10$^{-12}$ &1.85 $\times$ 10$^{-12}$ \\
\hline
2.226   &0.896  &1.52 $\times$ 10$^{-12}$ &4.62 $\times$ 10$^{-13}$ \\
\hline
3.320   &1.337  &5.79 $\times$ 10$^{-13}$ &1.92 $\times$ 10$^{-13}$ \\
\hline
4.953   &1.994  &2.01 $\times$ 10$^{-13}$ &9.20 $\times$ 10$^{-14}$ \\
\hline
7.389   &2.975  &7.73 $\times$ 10$^{-14}$ &4.01 $\times$ 10$^{-14}$ \\
\hline
11.023  &4.439  &1.63 $\times$ 10$^{-14}$ &1.92 $\times$ 10$^{-14}$ \\
\hline
\end{tabular} 
\end{center}
\end{table}

\begin{figure}
\begin{center}
\includegraphics[width=14cm,angle=0,clip]{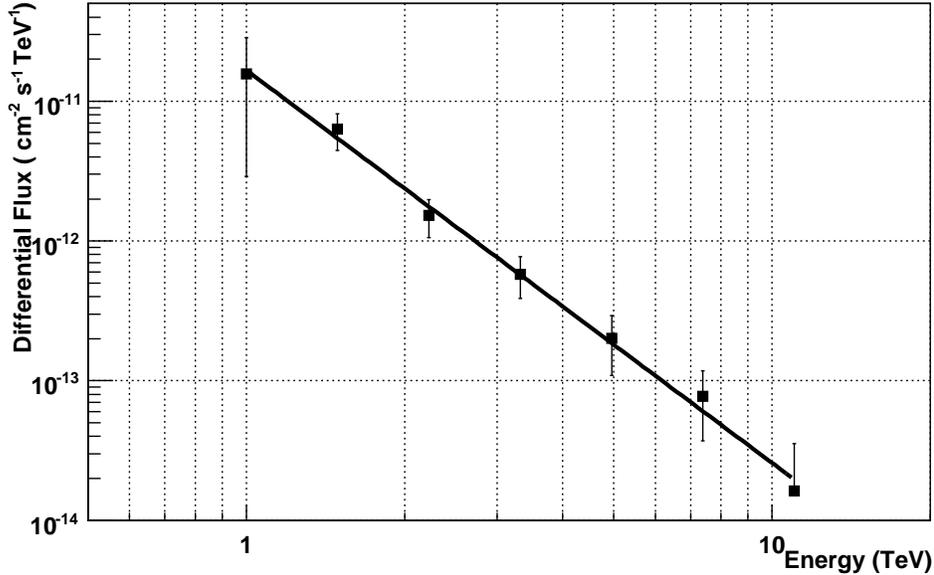}
\caption{Differential energy spectrum of Mrk 501 obtained  during 2006 observations.   Error bars shown  are for statistical errors only.} \label{fig:501sp}
\end{center}
\end{figure}

The  differential energy  spectrum obtained  after  applying the Dynamic Supercuts to the combined 2006  data is shown in Fig. \ref{fig:501sp}.  A power law fit to the data ($d\Phi/dE=f_0 E^{-\Gamma}$) in the energy range 1-11 TeV yields $f_0=(1.66\pm0.52)\times 10^{-11}cm^{-2}s^{-1}TeV^{-1}$ and  $\Gamma=2.80\pm0.27$ with a $\chi^2/dof= 0.773/5$ (probability =0.97). Again a low value of the reduced  $\chi^2$ may be due to overestimation of the errors in the present measurements. The errors in the flux constant and   spectral index  are   standard errors.  
We have also tried a  power law with an exponential cutoff function of the form ($d\Phi/dE=f_0E^{-\Gamma}exp(-E/E_0)$ while fitting the observed source differential
energy spectrum  which, however,  does not indicate  any  exponential cutoff feature in the energy range 1-11 TeV of the source spectrum.   One of the possible reasons  for not having obtained exponential cutoff  in the reported source  energy spectrum could be the low statistics ( about 500  $\gamma$-ray like events only) detected  with the TACTIC telescope.       

\section{Discussion and Conclusions}
We have  studied  the BL Lac object Mrk 501  in  
VHE gamma-ray energy range   with the TACTIC  $\gamma$-ray    
telescope during 2005- 06. We do not find any evidence for the presence of a statistically significant VHE gamma-ray
signal  either in the overall data or   when the data is analysed       on the month to month basis  or a night to night basis  during 2005 observations. An upper limit of I($\geq$1 TeV)$\leq 4.62 \times10^{-12}$ photons cm$^{-2}$  s$^{-1}$ ( 28$\%$ of the TACTIC detected Crab Nebula flux) has been placed  at a 3$\sigma$ confidence level on the integrated $\gamma$-ray flux.
% and we conclude that the source was below the resulted upper limit flux value  during 2005 TACTIC observations.
Our results do not conflict with those of the MAGIC group \cite{magic07} in 2005 as the two  observations are not overlapping in time as mentioned earlier. These observations rather provide the source  light curve prior to the MAGIC detected two historical flares on June, 30 and July, 9, 2005.            
\begin{figure}
\begin{center}
\includegraphics[width=14cm,angle=0,clip]{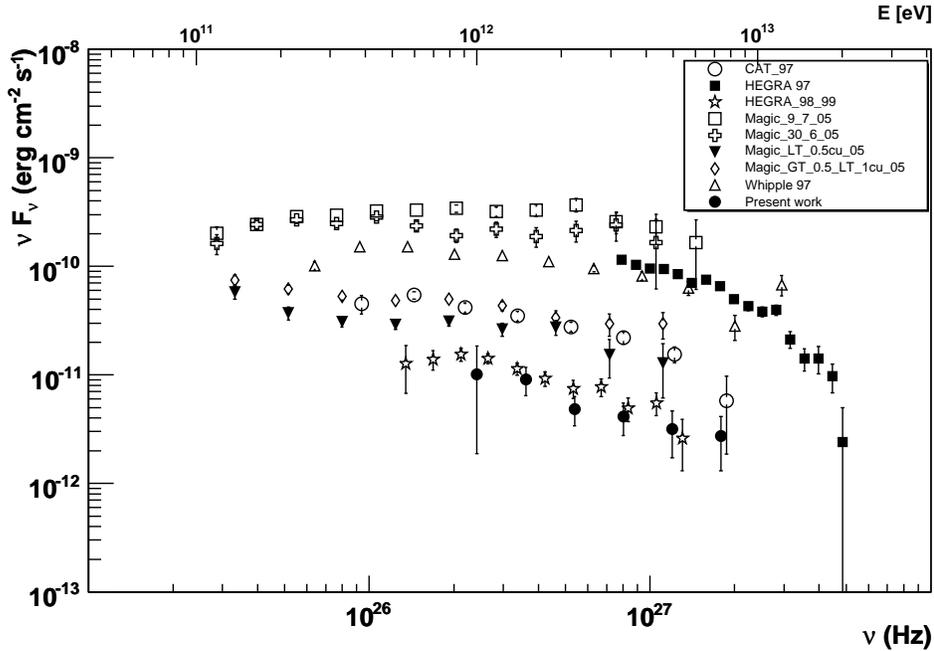}
\caption{Comparison of VHE  TACTIC derived Mrk 501 SED with  those  obtained during different epochs  by  WHIPPLE \cite{Samuelson98}, HEGRA \cite{Aharonian01,Aharonian01a} , CAT\cite{cat99}, MAGIC\cite{magic07} collaborations.} \label{fig:comp}
\end{center}
\end{figure}

During 2006 observations we have detected a TeV  $\gamma$-ray signal from the source direction  with a statistical significance of 7.5$\sigma$ in 66.8 hours of on-source observations.  Monthly spellwise analysis revealed that the source was relatively  brighter during  II spell of  2006 observations as its statistical significance is close to 6$\sigma$ in 27.71 hours of observations. 
%Results of the data analysis,   on nightly basis, in order  to explore the possibility of  a very strong episodic TeV emission which is a common feature such objects, are shown in  Figure \ref{fig:lc}b. 
Data collected during three  nights of March 26, 27 and 28, 2006 ( MJD 53820, 53821 and  53822) do reveal   hourly gamma-ray rate more than two times  that of the Crab Nebula.  Excesses of gamma-ray like events obtained are 37.0 $\pm$ 12.7, 41.7 $\pm$ 11.7 and   56.0 $\pm$ 15.5   for 1.9, 1.7 and 2.8 hours of observations  respectively. These  observations possibly  suggest that the source has  gone into a   variable active  state in the VHE region during 
the observations of
aforementioned  three nights. We have  
measured  the  source differential energy spectrum
which fits well with  the power law function of the form  $d\Phi/dE=f_0 E^{-\Gamma}$ with $f_0=(1.66\pm0.52)\times 10^{-11}cm^{-2}s^{-1}TeV^{-1}$ and  $\Gamma=2.80\pm0.27$ with a $\chi^2/dof= 0.773/5$ (probability =0.97) in the energy range 1-11 TeV. We do  not  get any evidence for  an  exponential cutoff like feature when we use   a  power law function   of the form $d\Phi/dE=f_0E^{-\Gamma}exp(-E/E_0)$ while fitting the observed source differential spectrum in the aforementioned TACTIC energy range. 

In Figure   \ref{fig:comp}  we compare important source related  VHE spectra results  obtained during different epochs  by various groups,  WHIPPLE \cite{Samuelson98}, HEGRA \cite{Aharonian01,Aharonian01a}, CAT\cite{cat99} and  MAGIC\cite{magic07}. As is clear from this figure,  TACTIC 2006 results   closely follow those  obtained by the HEGRA group during the period 1998-99, except for the exponential cutoff feature with $E_0$ =2.61 TeV.  
This source along with  Mrk 421 is very interesting as it is the second nearest BL Lac object,  which has  been observed in highly variable   states by aforementioned groups.  In order to understand such highly violent extra galactic objects  it is necessary  to continue monitoring them for long  durations with sensitive   $\gamma$-ray telescopes.               
\section{Acknowledgements}
The authors would like to convey their gratitude to all the concerned colleagues of the Astrophysical Sciences Division for their contributions towards the instrumentation and observation aspects of the TACTIC telescope.  We are  thankful to Dr. D. Paneque of Max-Planck-Institut f\"ur Physik,  M\"unchen, Germany,  for providing  useful inputs with respect to  the MAGIC Mrk 501 work.   We  also  thank the anonymous referees for their useful comments.

\end{document}